
\input phyzzx
%



\let\lettertopfil=\lettertopskip




\def\l"{``} 
\def\linebreak{\unskip\break}

%
%

\def\ifmath#1{\relax\ifmmode #1\else $#1$\fi}

\def\unlock{\catcode`@=11} 
\def\lock{\catcode`@=12} 



%

%

%

%

\def\uss#1{\vtop{\hbox{#1}\kern 3pt \hrule}} 
\def\~{\accent'24 } 

\def\ls#1{\ifmath{_{\lower1.5pt\hbox{$\scriptstyle #1$}}}}

\def\etal{{\it et al.}}%
\def\ie{{\it i.e., }}%

\def\SLACHEAD{\setbox0=\vbox{\baselineskip=12pt
      \ialign{\tenfib ##\hfil\cr
         \hskip -17pt\tenit Mail Address:\ \ Bin 81\cr
         SLAC, P.O.Box 4349\cr Stanford, California, 94305\cr}}
   \setbox2=\vbox to\ht0{\vfil\hbox{\caps Stanford Linear Accelerator
         Center}\vfil}
   \smallskip \line{\hskip -7pt\box2\hfil\box0}\bigskip}
%
%
\def\SCIPP{\centerline {\it Santa Cruz Institute for Particle Physics}
  \centerline{\it University of California, Santa Cruz, CA 95064}}

\newbox\figbox
\newdimen\zero  \zero=0pt
\newdimen\figmove
\newdimen\figwidth
\newdimen\figheight
\newdimen\figrefheight
\newdimen\textwidth
\newtoks\figtoks
\newcount\figcounta
\newcount\figlines
\def\figreset{\global\figmove=\baselineskip \global\figcounta=0
\global\figlines=1 \global\figtoks={ } }


\def\picture#1by#2:#3{\global\setbox\figbox=\vbox{\vskip #1
\hbox{\vbox{\hsize=#2 \noindent #3}}}
\global\setbox\figbox=\vbox{\kern 10pt
\hbox{\kern 5pt \box\figbox \kern 10pt }\kern 10pt}
\global\figwidth=1\wd\figbox
\global\figheight=1\ht\figbox
\global\figrefheight=\figheight
\global\textwidth=\hsize
\global\advance\textwidth by - \figwidth }
\def\figtoksappend{\edef\temp##1{\global\figtoks=%
{\the\figtoks ##1}}\temp}
\def\figparmsa#1{\loop \global\advance\figcounta by 1%
\ifnum \figcounta < #1 \figtoksappend{0pt \the\hsize}%
\global\advance\figlines by 1 \repeat }
\newdimen\figstep
\def\figst@p{\global\figstep = \baselineskip}
\def\figparmsb{\loop \ifdim\figrefheight > 0pt%
\figtoksappend{ \the\figwidth \the\textwidth}%
\global\advance\figrefheight by -\figstep%
\global\advance\figlines by 1%
\repeat }


\def\figtext#1:#2{\figreset \figst@p%
\figparmsa{#1}%
\figparmsb%
\multiply\figmove by #1%
\global\setbox\figbox=\vbox to 0pt{\vskip\figmove\hbox{\box\figbox}\vss}%
\parshape=\the\figlines\the\figtoks\the\zero\the\hsize%
\noindent\rlap{\box\figbox} #2}

\def\lozenge{\boxit{\hbox to 1.5pt{\vrule height 1pt width 0pt \hfill}}}

\def\cornerarrow{%
   \rlap{ 
      \raise 7pt
      \hbox{\vrule height 6.5pt depth 4pt}%
      }%
   \hskip 0pt plus 10000pt \rightarrow
   }%
%
%

\def\9{\hskip .5em}

\def\|{\vrule height 16pt depth 6 pt}

\def\spose#1{\raise6pt\hbox to 0pt{#1 \hskip0pt minus10000pt}}
\def\.{\hskip -4pt plus 10000pt}

\def\leftline#1{\line{#1\hss}}

\def\rightline#1{\line{\hss#1}}

\def\mybox#1{\hbox to size{\vbox{\hrule\hbox{\vrule\hskip4pt
\vbox{\vskip4pt #1 \vskip4pt}\hskip4pt\vrule}\hrule}}}

\def\lead{\leaders\hbox to 10pt{\hfill.\hfill}\hfill}

\chapterminspace=12pc
\referenceminspace=12pc
\def\refout{\par\penalty-400\vskip\chapterskip
   \spacecheck\referenceminspace
   \ifreferenceopen \Closeout\referencewrite \referenceopenfalse \fi
   \line{\hfil REFERENCES\hfil}\vskip\headskip
   \input \jobname.refs
   }
%
%
\def\PRL#1&#2&#3&{\sl Phys.~Rev.~Lett.\ \bf #1\rm ,\ #2\ (19#3)}
\def\PRB#1&#2&#3&{\sl Phys.~Rev.\ \bf #1\rm ,\ #2\ (19#3)}
\def\NPB#1&#2&#3&{\sl Nucl.~Phys.\ \bf #1\rm ,\ #2\ (19#3)}
\def\PL#1&#2&#3&{\sl Phys.~Lett.\ \bf #1\rm ,\ #2\ (19#3)}
\def\ZP#1&#2&#3&{\sl Z.~Phys.\ \bf #1\rm ,\ #2\ (19#3)}
\def\NIM#1&#2&#3&{\sl Nucl.~Instrum.~\& Methods\ \bf #1\rm ,\ #2\ (19#3)}
\def\RMP#1&#2&#3&{\sl Rev.~Mod.~Phys.\ \bf #1\rm ,\ #2\ (19#3)}
\def\IEEE#1&#2&#3&{\sl IEEE Trans.~Nucl.~Phys. \bf #1\rm ,\ #2\ (19#3)}
\def\NPRL#1&#2&#3&{\sl Phys.~Rev.~Lett.\ \bf #1\ \rm (19#2)\ #3}
\def\NPR#1&#2&#3&{\sl Phys.~Rev.\ \bf #1\ \rm (19#2)\ #3}
\def\NNP#1&#2&#3&{\sl Nucl.~Phys.\ \bf #1\ \rm (19#2)\ #3}
\def\NPL#1&#2&#3&{\sl Phys.~Lett.\ \bf #1\ \rm (19#2)\ #3}
\def\NZP#1&#2&#3&{\sl Z.~Phys.\ \bf #1\ \rm (19#2)\ #3}
\def\AJ#1&#2&#3&{\sl Ap.~J.\ \bf #1\ \rm (19#2)\ #3}
%
%
\def\CVaddresses#1#2{\hbox to\hsize{
Home:\quad\vtop{\halign{##\hfil\crcr #1}}\hfill
Work:\quad\vtop{\halign{##\hfil\crcr #2}}}}
%
%
\def\pmb#1{\setbox0=\hbox{$#1$}%
     \kern-.025em\copy0\kern-\wd0
     \kern.05em\copy0\kern-\wd0
     \kern-.025em\raise.0433em\box0 }

\def\spmb#1{\setbox0=\hbox{$\tenpoint #1$}%
     \kern-.020em\copy0\kern-\wd0
     \kern.03em\copy0\kern-\wd0
     \kern-.020em\raise.0433em\box0 }
%

%

%
%
%
\unlock                         
\def\addressee#1{\line{\hskip10cm\the\date\hfill} \bigskip
   \vskip\lettertopfil
   \ialign to\hsize{\strut ##\hfil\tabskip 0pt plus \hsize \cr #1\crcr}
   \writelabel{#1}\medskip \noindent\hskip -\spaceskip \ignorespaces }
\def\signed#1{\par \penalty 9000 \bigskip \dt@pfalse
  \everycr={\noalign{\ifdt@p\vskip\signatureskip\global\dt@pfalse\fi}}
  \setbox0=\vbox{\singlespace \halign{\tabskip 0pt \strut ##\hfil\cr
   \noalign{\global\dt@ptrue}#1\crcr}}
  \line{\hskip10cm \box0\hfil} \medskip }




\overfullrule=0pt

%
%
\def\slacletters{\letters \let\letterhead=\SLACHEAD }
\def\ucscletters{\letters \let\letterhead=\UCSCHEAD }
\def\myucletters{\letters \let\letterhead=\MYUCHEAD }
\def\SLACHEAD{\setbox0=\vtop{\baselineskip=10pt
     \ialign{\eightrm ##\hfil\cr
        \slacbin\cr
        P.~O.~Box 4349\cr
        Stanford, CA 94309\cropen{1\jot}
        \slacphone\cr }}%
   \setbox2=\hbox{\caps Stanford Linear Accelerator Center}%
   \hrule height \z@ \kern -0.5in
   \vbox to 0pt{\vss\centerline{\seventeenrm STANFORD UNIVERSITY}}
   \vbox{} \medskip
   \line{\hbox to 0.7\hsize{\hss \lower 10pt \box2 \hfill }\hfil
         \hbox to 0.25\hsize{\box0 \hfil }}\medskip }
\def\UCSCHEAD{\setbox0=\vtop{\baselineskip=10pt
     \ialign{\eightrm ##\hfil\cr
        SCIPP\cr Natural Sciences II\cr
        University of California\cr
        Santa Cruz, CA 95064\cropen{1\jot}\cr}}
   \setbox2=\hbox{\caps Santa Cruz Institute for Particle Physics}%
   \vbox to 0pt{\vss\centerline{\seventeenrm UNIVERSITY of CALIFORNIA}}
   \vbox{} \medskip
   \line{\hbox to 0.7\hsize{\hss \lower 10pt \box2 \hfill }\hfil
         \hbox to 0.25\hsize{\box0 \hfil }}\medskip }
\def\MYUCHEAD{\rightline{\vbox{\baselineskip=10pt
         \hrule height 0pt width 0.25\hsize
         \ialign{\eightrm ##\hfil\cr
              Prof. Michael Dine\cr
              Physics Dept.\cr
              Natural Sciences II\cr
              University of California\cr
              Santa Cruz, CA  95064\cropen{1\jot}
              (408) 459--3033\cr
        Bitnet:  DINE@SLACVM\cr}}}
    \medskip }
\lock

\normalparskip=0pt
\parindent 36pt

\unlock
\def\refitem#1{\r@fitem{#1.}}
\def\chapter#1{\par \penalty-300 \vskip\chapterskip
   \spacecheck\chapterminspace
   \chapterreset \leftline{
\bf \chapterlabel.~#1}
   \nobreak\vskip\headskip \penalty 30000
   {\pr@tect\wlog{\string\chapter\space \chapterlabel}} }
\lock
\def\twothirds{{\textstyle{2 \over 3}}}
\def\SCIPP{\centerline{\it Santa Cruz Institute for Particle Physics}
  \centerline{\it University of California, Santa Cruz, CA 95064}}
\Pubnum{SCIPP 92/21}
\date{June 1992}
\pubtype{ T}
\titlepage
\vskip3cm
\singlespace
\centerline{ELECTROWEAK BARYOGENESIS:}
\centerline{AN OVERVIEW (WHERE ARE WE NOW?)}
\foot{Work supported in part by the U.S. Department of Energy.}
\vskip12pt
\centerline{\caps Michael Dine}
\vskip2pt
\SCIPP
\vfill
\vbox{ \narrower
\centerline{ABSTRACT}

It now seems plausible that the observed baryon asymmetry may have
been produced at the electroweak phase transition.  We review
the considerations which lead to this conclusion,
focusing on the obstacles to making reliable
estimates.  These arise from incomplete knowledge of baryon-violating
rates near the phase transition, our limited understanding of the
phase transition itself, and incomplete understanding of the
mechanisms which actually lead to an asymmetry.}

\vfill
\centerline {Invited Lecture}
\centerline{Texas Symposium on Electroweak Baryon Number Violation}
\centerline{Yale University, March 1992.}
\vfill
\endpage

\chapter{Introduction}

\REF\shaposh{M.E. Shaposhnikov,  Phys.~Lett. {\bf 277B} (1992) 324.}
In principle, the
three conditions for baryogenesis enunciated by Sakharov can all be met by
electroweak physics.  Baryon number is badly violated by the electroweak
theory at temperatures at or above the electroweak phase transition.
The weak phase transition itself can be first order, providing a
departure from equilibrium.  Substantial CP violation can be provided by
extensions of the minimal standard model, such as supersymmetric and
multi-Higgs theories.\foot{Shaposhnikov has made a number of exotic
suggestions as to how KM phases might yield a large enough asymmetry.
See, for example, reference \shaposh.}  Indeed, provided that the transition
is actually first order, it is clear that one will necessarily produce
some baryon number at the transition.  In this talk I will attempt to
survey our current ability to calculate the asymmetry which arises
in any given extension of the standard model.  Corresponding to
Sakharov's three
conditions, this problem has three aspects:  computing the rate
for baryon number violation in different environments; determining the
nature of the phase transition as a function of the underlying theory;
understanding precisely which non-equilibrium processes are most
important to generating the asymmetry, and computing their contributions.

Before discussing detailed analyses, we should first ask:  why
is the prospect of electroweak baryogenesis interesting?  There are
at least two answers.  First, the ties between cosmology and
experimentally accessible particle physics are very limited.
It would be very satisfying if at some stage we could
use experimental measurements of microscopic parameters to
predict or explain the abundance of matter.  It would be even
more dramatic if, by insisting that the baryon asymmetry be of
electroweak origin, we could make predictions for future
particle experiments.  Second, there is a more theoretical
concern, which is that such late production of the asymmetry
may ameliorate some of the problems of inflationary cosmology,
particularly problems connected with reheating.
On the other hand, electroweak
baryogenesis is unlikely to leave observable astrophysical
signals (apart from matter itself).   At this temperature, for example,
there are only about $10^{17}$ baryons within a horizon volume.

\bigskip\noindent
{\bf 2.~Baryon Number Violation in the Standard Model at High}\nextline
{\bf Temperature}
\vskip8pt
\REF\linde{A. D. Linde, Phys.~Lett. {\bf 70B} (1977) 306;}
\REF\dimopoulos{S. Dimopoulos and L. Susskind, Phys.~Rev. {\bf D18}
(1978) 4500.}
\REF\manton{N. Manton, Phys.~Rev. {\bf D28} (1983) 2019;
D. Klinkhammer and N. Manton,
 Phys.~Rev. {\bf D30} (1984) 2212.}
\REF\krs{V.A. Kuzmin, V.A. Rubakov and M.E. Shaposhnikov,
Phys.~Lett. {\bf B155} (1985) 36.}
The possibility that effects associated with the anomaly might
be enhanced at high temperature, and that this might have importance
for the baryon asymmetry was suggested early on by Linde\refmark{\linde}
and by Dimopoulos and Susskind.\refmark{\dimopoulos}  Klinkhammer
and Manton, in their original work on sphalerons, argued that
B violation would be enhanced at high temperature (but not at high
energies).\refmark{\manton}  However, the first to really explore
this possibility in a serious way were Kuzmin, Rubakov and
Shaposhnikov.\refmark{\krs}  These authors estimated the rates
for B violation in the broken phase of the theory, and considered
possible ramifications.  In particular, they noted that these effects
could modify or wipe out any pre-existing baryon number, and that
the electroweak transition itself might be the origin of the baryon
number if it is first order.

The basic ideas underlying these phenomena are very simple.
While the lagrangian of the standard model preserves $B$ and $L$
at the classical level, both suffer from an anomaly, and are not
strictly conserved ($B-L$ is free of anomalies).  For example,
$$\partial_{\mu}j^{\mu~B} = {3 g_2^2 \over 16 \pi^2} F \tilde F.
\eqn\anomaly$$
$F \tilde F$ can itself be written as a total divergence,
$$F \tilde F = \partial_{\mu} K^{\mu}\eqn\kmu$$
where
$$K^{\mu}= \epsilon_{\mu \nu \rho \sigma} Tr(A_{\nu}\partial_{\rho}
A_{\sigma} + {\twothirds} A_{\nu} A_{\rho} A_{\sigma}).\eqn\kmuequals$$
The zero component of $K$ is known as the Chern-Simons number, $n_{CS}$.
This number plays a special role in gauge theories.  In a theory
without fermions, there are classically an infinite set of
zero energy states labeled by $n_{CS}$.
In the standard model,
as a consequence of the anomaly, states of different $n_{CS}$ differ
also in their baryon number.  Thus states of different baryon number
are separated from one another by continuous changes in bosonic
\REF\thooft{G. 't Hooft, Phys.~Rev.~Lett. {\bf 37} (1976) 8;
Phys.~Rev. {\bf D14} (1976) 3432.}
fields.  At low energies, one can pass from one such state to another
by tunneling.  The corresponding amplitude was computed by
't Hooft,\refmark{\thooft}
using instanton methods, and is -- as expected of a tunneling
amplitude in a weakly coupled theory -- exponentially small,
$$\Gamma \sim e^{-4 \pi / \alpha_W}. \eqn\zerotemp$$

\REF\banksetal{T. Banks, M. Dine, G. Farrar, D. Karabali and B.
Sakita, Nucl.~Phys. {\bf B347} (1990) 581;
M. Dine, Santa Cruz preprint SCIPP 90/27 (1990), in
Proceedings of 18th SLAC Summer Institute (1990).}
It is natural to ask whether the tunneling rate is enhanced at high
temperatures and energies.  The question of high energy enhancement
will be the subject of the second half of this meeting.  I have expressed
my views -- that the cross
section grows exponentially with energy for some range of energy, but
is always exponentially small --  on this controversial subject
elsewhere\refmark{\banksetal} and
will not elaborate them again here.  At high temperatures, however,
\REF\affleck{I. Affleck, Phys.~Rev.~Lett. {\bf 46} (1981) 388.}
it is now widely accepted that the rate becomes large.  This is
readily understood by considering an analogy with a system with
\FIG\onedof{Potential for analog system with a single degree of freedom.
Curvature in the metastable minimum is $\omega^2$; the barrier height
if $V_o$.}
a single degree of freedom (see fig.~\onedof).\refmark{\affleck}  The system
possesses a metastable
minimum, in which the curvature of
the potential is $\omega^2$; the barrier between this minimum and the true
minimum has height $V_o$.  For such a
system at zero temperature, in the classical limit one can perform
a WKB analysis, giving an exponential suppression.  For
temperatures large compare to the curvature in the metastable minimum,
$\omega^2$, one can apply classical statistical mechanics to see that the
barrier penetration rate goes as $e^{-\beta V_o}$.  The analogs
of the metastable minimum in the electroweak theory are states centered
around different integer values of $n_{CS}$.  The point at the top
of the barrier is the ``sphaleron,"\refmark{\manton} a static solution
of the classical equations of motion with a single unstable mode.
The barrier height is just the sphaleron energy,
$$E_{sp} = A M_W{T} /\alpha_W \eqn\sphaleronenergy$$
corresponding to a transition rate
$$\Gamma \propto e^{-A M_W /(\alpha_W T)}.\eqn\sphaleronrate$$

\FIG\instanton{Schematic drawing of the potential of a gauge
theory as a function of Chern-Simons number.
Instanton describes a  path in field space
from one vacuum to another; the highest point along this
path has energy $E_b$.}

\REF\arnold{P. Arnold and L. McLerran, Phys.~Rev. {\bf D37}
(1988) 1020.}
\REF\others{E. Mottola and A. Wipf, Phys.~Rev. {\bf D39} (1989)
588; S. Khlebnikov and M. Shaposhnikov,
Nucl.~Phys. {\bf B308} (1988) 885; M. Dine, O. Lechtenfeld,
B. Sakita, W. Fischler and J. Polchinski, Nucl.~Phys.
{\bf B342} (1990) 381.}
All of this is for temperatures below the critical temperature for the
electroweak phase transition, $T_c$.  For higher temperatures
$M_W=0$, and it is believed that
the rate behaves as\refmark{\arnold,\others}
$$\Gamma = \kappa(\alpha_W T)^4.
\eqn\hightemprate$$
This result can easily be understood in a heuristic way as follows.
The high temperature theory is essentially an unbroken gauge
theory in  $2+1$ dimensions.   It is believed that such a system
has a correlation length (``magnetic mass" or ``magnetic screening length")
$$\zeta \sim (\alpha_W T)^{-1}.
\eqn\correlation$$
Return to the zero temperature, four dimensional theory, and
consider a pure gauge instanton with scale size $\rho$.  Such an instanton
can be viewed as parameterizing a path in field space starting
in a state with one value of $n_{CS}$ and ending in another,
as the ``imaginary time" varies from $-\infty$ to $\infty$.
(see fig.~\instanton).
For a given value of $\rho$, the highest
energy point along this path has energy $E_b$, where, on
dimensional grounds, $E_b \sim {1/(\alpha_W \rho)}$.
The largest allowed $\rho$ gives the smallest barrier.  Since
one clearly can't have $\rho \gg \zeta$, the trajectory with
the lowest possible barrier has $E_b \sim T$.  As a result, there
is no exponential suppression of the rate.  The rest of eq.~%
\hightemprate\ follows on dimensional grounds.
In their earlier work on this subject, Arnold and McLerran
found, from heuristic considerations, that $\kappa$ could be expected
to be of order $10$ or so.\refmark{\arnold}

This whole picture has received support from a number of sources.
Arnold and McLerran\refmark{\arnold}
were among the first to take the suggestion
of Kuzmin \etal seriously.  In their work they  addressed many
issues, including the problem of calculating the prefactor
in eq.~\sphaleronrate.  By studying a number of illustrative
toy models, they were also able to resolve a number of puzzles.  In particular,
they understood how one could have large transition rates while
at the same time having exponentially small $\theta$-dependence.
Skepticism was further damped by the numerical work of
Grigoriev, Rubakov and Shaposhnikov on the Abelian Higgs model
in $1+1$ dimensions.  At finite volume, this theory has a structure
very similar to four dimensional Yang Mills theory.  Classical
zero energy states are solutions of the equation
$$D_1 \phi = 0.\eqn\phieqn$$
With periodic boundary conditions, these are of the form
$$\phi = \phi_o e^{2 \pi i n/L}~~~~~~~~~~~~~
A_1 = {2 \pi n/L}.$$
The Chern-Simons number in this theory is very simple:
$$n_{CS} = {1 \over 2 \pi} \int dx A_1 = n. \eqn\onedcs$$
This theory admits sphalerons and instantons quite analogous
to those in $3+1$ dimensions, and one can estimate the transition
rate as discussed in refs. \krs\ and \arnold.

\REF\grigorievetal{D. Grigoriev, V. Rubakov and
M. Shaposhnikov, Phys.~Lett. {\bf 216B} (1989) 172.}
\FIG\grigoriev{Figure from ref. \grigorievetal, showing
discrete jumps in the Chern-Simons number.}
\FIG\unwind{Detail of one of the jumps.  Note that the scalar
field unwinds, and develops a zero somewhere.}

In order to verify if this picture is really correct, Grigoriev
\etal\ simulated the system numerically.  The basic idea in their
approach is to note first that at high temperatures the system
should be described by classical statistical mechanics.
Moreover, for sufficiently large volume, the evolution of a given
classical configuration should be ergodic, given that the system
is non-linear.  Thus they picked field configurations from
a statistically generated ensemble, and simply followed their
evolution for long periods of time.  As one can see in
fig.~\grigoriev,
the results are quite striking.  $n_{CS}$ makes small, rapid oscillations
about integer values for some period of time, and then less frequently
makes integer jumps.  The rates for these integer jumps are consistent
with sphaleron estimates.  Moreover, if one follows the scalar field
during one of these jumps, one sees that it ``unwinds" precisely
as one expects (see fig.~\unwind;
note that
the field develops a zero as it must
for non-singular evolution).  Closer study of these results reveals
many satisfying features.  The rate and amplitude of the small
oscillations, for example, can easily be estimated and is consistent
with the numerical results.  Overall, the picture is quite compelling.

\REF\carsonetal{L. Carson, Xu Li, L. McLerran and R.-T. Wang,
Phys.~Rev. {\bf D42} (1990) 2127.}
Further progress has also been made in understanding the four dimensional
situation.  In particular, Carson, Li, McLerran and Wang\refmark{\carsonetal}
have evaluated
the functional determinant about the sphaleron, which determines
the prefactor.  For example, for a single Higgs with self coupling
of order one, the rate is given by
$$\Gamma = \gamma \alpha_2^{-3} M_W^7 e^{-E_{sp}/T}.\eqn\fullrate$$

\REF\ambjornetal{J. Ambjorn, T. Askgaard, H. Porter and M.E.
Shaposhnikov, Nucl.~Phys. {\bf B353} (1991) 346.}
Simulations in four dimensions, of course, are more difficult.
It is particularly important, as we will see, to determine the rate
in the high temperature phase.  Ambjorn \etal\ have attempted
to simulate the high temperature rate in a manner similar to
the $1+1$ calculations.\refmark{\ambjornetal}  Again one observes
that $n_{CS}$ makes discrete jumps; the simulations yield
values for $\kappa$ in the range 0.1--1.0.
 While the results are impressive,
they are (not surprisingly) not quite as convincing as the one dimensional
calculations.  They are, for example, a good deal noisier.  This results,
in part, from the fact that if one simply computes the correlation
function for $n_{CS}$ in perturbation theory, it exhibits both
infrared and ultraviolet sensitivity.  Moreover, these authors are
not able to observe the expected $\sqrt{t}$ growth in $n_{CS}$.
Finally, the interpretation of their results is complicated
by the fact that they considered a theory with Higgs fields.
Improved simulations, both with and without Higgs fields, are very
important, since, as we will see, the baryon
asymmetry which arises in any model is roughly proportional to $\kappa$.
Heuristic models such as that of ref.~\arnold\ and one to
be discussed below suggest that $\kappa$ could be as large as $10$ to
$100$.
\chapternumber=2
\chapter{The Phase Transition in the Minimal Standard Model}

One might expect that the weak phase transition in the
minimal standard model would be completely
understood by
\REF\kirshnitz{D.A. Kirzhnits, JETP Lett. {\bf 15} (1972) 529;
D.A. Kirszhnits  and A.D. Linde, Phys.~Lett. {\bf 42B} (1972) 471.}
now, given that the subject has been studied for over
$20$ years, and was indeed the first setting in which phase transitions
in relativistic quantum field theories were considered.\refmark{\kirshnitz}
Yet surprisingly, many questions which turn out to be of great
importance for computing the baryon asymmetry remain unsettled.
In the literature, one finds what we might refer to as the ``usual"
picture of the phase transition:  the gauge symmetry is
\REF\lindebook{A..D. Linde, {\it Particle Physics and Inflationary
Cosmology} (Harwood, New York, 1990).}
unbroken at low $T$, unbroken for large $T$.\refmark{\kirshnitz,\lindebook}
To derive this
picture, one computes the (finite temperature) effective potential,
$V(\phi,T)$,
at one loop.  Since a theory with a single light Higgs doublet
already exhibits a first order phase transition, we will consider
this case first.
Contributions of particles of
mass $m(\phi)$ to $V(\phi,T)$ are proportional to $m^2\,T^2$, $m^3
\,T$ and
$m^4 \ln (m/T)$. We will assume that the Higgs boson mass
is smaller than the masses of W and Z bosons and the
top quark, $m_H < m_W, m_Z, m_t$. Therefore we will
neglect the Higgs boson contribution to $V(\phi,T)$.

The zero temperature potential, taking into account
one-loop corrections, is given by\refmark{\lindebook}
$$
V_0 = - {\mu^2\over 2}\phi^2 + {\lambda\over 4} \phi^4 +
2Bv_o^2\phi^2 - {3\over 2} B\phi^4 + B \phi^4 \ln\left({\phi^2\over
v_o^2}\right) \ .
\eqn\zerotemp$$
Here
$$
B = {3\over 64 \pi^2 v_o^4} (2 m_W^4 + m_Z^4 - 4 m_t^4) \ ,
\eqn\bdefined$$
$v_o = 246$ GeV is the value of the scalar field at the minimum
of $V_0$, $\lambda = \mu^2/v_o^2$, $m^2_H = 2\mu^2$.
At a finite temperature, one should add to this expression the term
$$
V_T = {T^4\over 2 \pi^2} \left[ 6I_{-}(y_W) + 3I_{-}(y_Z) -
6I_{+}(y_t)\right] \ ,
\eqn\tempcorrection$$
where $y_i = M_i\phi/v_o T$, and
$$
I_{\mp}(y) = \pm \int_{0}^{\infty} dx
\ x^2 \ln \left(1  \mp e^{- \sqrt{x^2+y^2}}\right) \ .
\eqn\iminus$$
In the high temperature limit it is sufficient to use an
\REF\ha{G. Anderson and L. Hall, Phys.~Rev. {\bf D45} (1992) 2685.}
approximate expression for $V(\phi,T)$\ \refmark{\kirshnitz,\ha}
$$
V(\phi,T) = D (T^2 - T_o^2) \phi^2 - E T \phi^3 +
{\lambda_T\over 4} \phi^4 \ .
\eqn\hightemp$$
Here
$$
D = {1\over 8v_o^2} ( 2 m_W^2 + m_Z^2 + 2 m_t^2) \ ,
\eqn\dequals$$
$$
E =  {1\over 4\pi v_o^3} ( 2 m_W^3 + m_Z^3) \sim 10^{-2} \ ,
\eqn\eequals$$
$$
T^2_o = {1\over 2D}(\mu^2 - 4Bv_o^2) =
{1\over 4D}(m_H^2 - 8Bv_o^2) \ ,
\eqn\tzero$$
$$
\lambda_T = \lambda - {3\over 16 \pi^2 v_o^4}
\left( 2 m_W^4 \ln{m^2_W\over a_B T^2} +
m_Z^4 \ln{m^2_Z\over a_B T^2} -
4 m_t^4 \ln{m^2_t\over a_F T^2}\right) \ ,
\eqn\lambdat$$
where $\ln a_B = 2 \ln 4\pi - 2\gamma \simeq 3.91$,
$\ln  a_F = 2 \ln \pi - 2\gamma \simeq 1.14$.

\REF\sher{M. Sher, Phys.~Rep. {\bf 179} (1989) 273.}

This potential leads to a  phase transition which is at least weakly
first order, basically as a consequence of the term cubic in $\phi$.
At very high temperatures, $V$ has a unique minimum at $\phi=0$.
As the temperature is decreased, a second minimum appears
at a temperature
$$T_1 = {T_0 \over 1 - 9 E^2/8 \lambda_T D}.\eqn\tone$$
At a temperature $T_c$, this new minimum becomes degenerate
with the minimum at the origin; $T_c$ and the corresponding minimum
of the potential $\phi_c$ are given by
$$T_c^2={T_o^2 \over 1 - E^2/\lambda D}~~~~~~~~~~\phi_c
=2E T_C /\lambda_{T_c}. \eqn\tc$$
Finally, at $T_0$, the minimum at the origin disappears, and the potential
has a unique minimum.

\REF\dhss{M. Dine, P. Huet, R. Singleton and L. Susskind, Phys.~Lett.
{\bf 257B} (1991) 351.}
\REF\mvst{L. McLerran, M. Shaposhnikov N. Turok and M. Voloshin,
Phys.~Lett. {\bf 256B} (1991) 451.}
\REF\bubble{A.D. Linde, Phys.~Lett. {\bf 70B} (1977) 306;
{\bf 100B} (1981) 37; Nucl.~Phys. {\bf B216} (1983) 421.}
One might expect then that the phase transition would proceed
through the formation of bubbles at some temperature between
$T_c$ and $T_0$,\refmark{\dhss,\mvst} and that the study of the
evolution of these bubbles would be straightforward.  In
the standard approach to the computation of finite temperature
nucleation rates\refmark{\bubble} one determines the structure
of the critical bubble by looking for stationary points of the
classical action which tend to zero at
large $\vert \vec x \vert$ and some non-zero
constant for small $\vert \vec x \vert$;
the rate of bubble production is then
given by the associated Boltzmann factor, $e^{-S_{cl} \beta}$.
As one approaches $T_c$ from below, $S_{cl}$ tends to $\infty$.
The expansion rate at this time is very small, so
at some slightly lower temperature
($T = .98 T_c$ or so), when the action is still quite
large ($S_{cl} \sim 140$) the universe
fills with bubbles. Early work of Linde\refmark{\bubble}
suggests that the
bubbles will be rather slow and thick in the case that the weak
transition is weakly first order, as it is for a Higgs with mass
comparable to the present LEP limits.

While this picture may be simple and appealing,
there are many unsettled issues and a good deal of controversy
surrounding almost all aspects of the phase transition.
Indeed, questions have been raised about the order and
nature of the phase transition, and, assuming that the phase
transition is first order, widely disparate results have been obtained
for the properties of the bubbles.  The discussion which follows
will focus largely on the phase transition in the minimal theory.
As we will see, for a variety of reasons, this theory cannot
be relevant to electroweak baryogenesis; still, given its
simplicity, it is a useful prototype, and the analyses
described here can readily be extended to other theories.

\FIG\phifour{One loop diagram which gives rise to $\phi^3$ term
in the potential.}
\REF\brahm{D. Brahm and S. Hsu, Caltech preprints CALT-68-1705
and CALT-68-1762 (1991)
and talk at this meeting.}
\REF\higgslimit{M.E. Shaposhnikov, JETP Lett. {\bf 44} (1986) 465;
Nucl.~Phys. {\bf 287} (1987) 757; Nucl.~Phys. {\bf 299} (1988) 797;
A.I. Bochkarev, S. Yu. Khlebnikov and M.E. Shaposhnikov,
Nucl.~Phys. {\bf B329} (1990) 490.}
\REF\dhs{M. Dine, P. Huet, and R. Singleton
SCIPP 91/08 (1991).}
\REF\dhll{M. Dine, P. Huet, R. Leigh, A. Linde
and D. Linde, SLAC-PUB-5740 (1992);
SLAC-PUB-5741.}
Let us consider first the question of the order of the phase
transition.  We have noted that the $\phi^3$ term in the potential
is crucial to the first order nature found at one loop.
However, the presence of this $\phi^3$ term is itself an
indication of an infrared problem.  The expression for $V(\phi,T)$
[eq.~\tempcorrection] is formally a function of $\vert \phi \vert^2$.
However, the expansion in powers of $\phi^2$ breaks down at order
$\phi^4$.  In terms of Feynman graphs, the diagram
of fig.~\phifour\
diverges linearly (the zero frequency term in the corresponding sum
has a divergent momentum integral).  The existence of such a term
raises the question of what happens in higher orders.
Recently, Brahm and Hsu\refmark{\brahm} and
Shaposhnikov\refmark{\shaposh}
have investigated aspects of this question.  Their investigations
yielded puzzling -- and opposite -- conclusions.  Both found
terms of order $g^3 \phi$ in the potential at two-loop order.
However, their answers differed in sign:  Brahm and Hsu found a positive term,
which tends to render the transition second order, while
Shaposhnikov found a negative sign, which tends to make the transition
more strongly first order.  Clearly both results cannot be correct --
and in fact these authors have all conceded some errors.  However,
the errors turn out to be subtle, and this work points out
the need for an improved understanding of the perturbation expansion.
P. Huet, R. Leigh, A. Linde and I have pursued these questions
thoroughly.\refmark{\dhll}
We have shown that through two-loop order there are no
linear terms and that perturbation theory is under control.  However,
we have also shown that the coefficient of the cubic term
is reduced by higher order corrections to $2/3$ of its one-loop
value.  This tends to make the phase transition more weakly first order.
In particular, this observation plus the present limits on the Higgs
mass from LEP completely rule out models with a single Higgs doublet
(even with injections of additional CP violation) as the source of the
baryon asymmetry.  The problem (as originally stressed by
Shaposhnikov\refmark{\higgslimit})
is that for a Higgs of mass $55$ GeV or larger, the Higgs vev after
the phase transition is significantly less than $T$, and the
baryon violation rate is larger than the expansion rate.  Thus
any pre-existing asymmetry is wiped out.  These topics are discussed
at greater length in the talks by R. Leigh , A. Linde, D. Brahm and S. Hsu
at this meeting.

\REF\john{M. Dine and J. Bognasco, SCIPP preprint in preparation.}
While the perturbation expansion is in reasonable shape, in the sense
that there are no unmanageable infrared problems, one can ask whether
it is really a good guide, \ie, whether for the temperatures of interest
the higher order corrections are numerically small. It is not hard
to evaluate the leading two-loop corrections (involving strong and weak
couplings) to the quadratic term in the potential and these tend to be
of order $20\%$.\refmark{\john}

\REF\gleiser{M. Gleiser and E. Kolb, FNAL preprint
FERMILAB-PUB-91/305-A (1991).}
\REF\tetradis{N. Tetradis, DESY preprint DESY-91-151.}
Even granted the (corrected) one-loop potential, questions have been
raised about how the transition actually proceeds.
Kolb and Gleiser\refmark{\gleiser}
and Tetradis\refmark{\tetradis} have raised questions about
the standard picture of bubble formation.  The basic idea is that,
if the transition is very weakly first order, then, even above
$T_c$, fluctuations are so large that the system can not be said
to lie in one state or another, but rather the universe consists
of a sort of emulsion of the two phases.  This quasi-equilibrium of
the two phases, they argue, is maintained by ``subcritical bubbles,"
bubbles smaller that the critical bubbles obtained from the analysis
above.  Certainly it is true that if the transition is sufficiently
weakly first order, the true and false vacuum will be separated
by an amount which is small compared to typical fluctuations in
the fields.  Linde will argue, in his talk, that if one simply
examines the typical fluctuations in these modes, one finds that
for any transition which is sufficiently strongly first order
to generate the baryon asymmetry, the conventional analysis
is correct; fluctuations are just not so large.

\REF\turok{N. Turok, Princeton University preprint PUPT-91-1273.}
\REF\mt{B. Liu, L. McLerran and N. Turok, Minn. preprint
TPI-MIN-92-18 (1992).}
All of this suggests that for some range of parameters, the
phase transition in the minimal standard model -- and modest extensions
of it -- will be first order.  Thus there is hope to produce
an appreciable baryon asymmetry at the electroweak transition.
Any asymmetry will be produced near the walls of the expanding
bubbles, for it is here that  there are significant departures
from equilibrium.  A typical bubble will grow to a macroscopic
size before colliding with other bubbles.  Thus it is
necessary to understand how the bubbles propagate through the hot plasma.
Naively,
one expects that the wall will be slowed by  particles
such as top quarks, $W$'s and $Z$'s, which gain appreciable mass
in the broken phase.\refmark{\bubble}
If the wall is thin compared to typical
mean free paths, these particles will transfer a (velocity-dependent)
amount of momentum to the wall, yielding a non-relativistic
terminal velocity.  In simple models, however, the wall
tends to be thick; the velocity is likely
to be substantially larger but the analysis becomes more
complicated.  Indeed, Turok\refmark{\turok} argued at one stage that
the wall might become ultrarelativistic.  More recently,
McLerran and Turok\refmark{\mt} and Huet \etal,\refmark{\dhll}
have made estimates
yielding wall velocities varying from about $0.2$ to $0.8$.  (These are
discussed in Leigh's talk at this meeting).  These analyses also
yield values for the wall thickness: for a 35 GeV Higgs, for
example, one finds $\ell \sim 40\ T^{-1}$.
However, the study of the wall motion is
still a quite primitive stage, and these results may be incomplete,
in the sense that potentially important (possibly dominant) effects
have not been accounted for.

\REF\cknone{A. Cohen, D. Kaplan and A. Nelson, Phys.~Lett.
{\bf 245 B} (1990) 561.}
\REF\bks{Bochkarev, Kuzmin, Shaposhnikov, Phys.~Lett. {\bf 244B}
(1990) 275.}
\REF\tztwo{N. Turok and J. Zadrozny, Nucl.~Phys. {\bf B369} (1992) 729.}
To conclude this section, I would like to return to the question
of bounds on the Higgs mass mentioned
above.  Shaposhnikov\refmark{\higgslimit}
has stressed that one needs the transition to be sufficiently first order
that once $\phi$ settles to its minimum, $\Gamma(\phi) \ll H$; otherwise,
the baryon number which has been generated will disappear
(unless $B-L$ is not conserved in the process\refmark{\cknone}).
Shaposhnikov found a limit of about $45$ GeV.  Allowing for the
corrections to the potential which have been described above,
and making generous estimates of the uncertainties in the various
calculations,\refmark{\dhs}
one finds a limit between 33 and 41 GeV.
Thus one needs to find models with a more strongly first order
transition.  In multi-Higgs models, detailed analyses show that the
limits
are significantly weakened.\refmark{\bks,\tztwo}
These analyses are rather complicated, however,
and Hall and Anderson\refmark{\ha} have provided a much simpler
example which gives an ``existence proof" that Higgs with mass
over $100$ GeV could be responsible for baryogenesis.  The idea
is simply to take the single Higgs model, and add a light scalar
singlet, $S$, with couplings
$$V_S= M^2 S^* S + 2 \zeta^2 S^*SH^*H + \lambda_S (S^*S)^2.
\eqn\hall$$
This scalar has two effects:
\item{a.}  if it is sufficiently light -- in practice, $150$ GeV or so --
it increases the magnitude of the cubic term in the finite-temperature
potential;
\item{b.}  the presence of $S$ modifies (through loop effects)
the relation
between the Higgs mass and the quartic coupling.

Indeed, Hall and Anderson find that it is quite easy to accommodate
a $150$ GeV Higgs in this framework.   One needn't take this model
too seriously, but it does suggest that if the electroweak transition
is the origin of matter, the Higgs sector may exhibit significant  structure at
relatively low energies.

\chapter{Producing the Baryon Asymmetry}

We have now seen that the weak phase transition may well be first
order, and that baryon number is badly violated at high temperatures.
The minimal standard model violates CP, and most extensions of the
minimal model violate CP more strongly.  Thus provided that the
electroweak transition is indeed first order, we are pretty much
guaranteed to produce an asymmetry; the question is how
large is the asymmetry.  We would like to be in the position
that, given a particular model we could simply calculate
the baryon to photon ratio
as a function of the model parameters.  In this
section, I will describe some of the scenarios which have been
proposed for producing the asymmetry.  We will see that there
are a number of mechanisms which might plausibly produce an adequate
$n_B / n_{\gamma}$,
but that the calculations still have large uncertainties.

The simplest -- if perhaps least efficient -- situation arises
when the effective theory at the weak transition contains only
a single Higgs doublet.\refmark{\dhss}  In this case, apart from
the usual KM phases, various higher dimension CP-violating operators
may arise from integrating out physics at higher scales.  This physics
might be supersymmetry, some version of technicolor, a multi-Higgs
model, or something else.  The lowest dimension CP-violating
bosonic operator one can write in this case is
$${\cal L}_{CP}= {g^4\sin(\delta)
\over 64 \pi^2 M^2}\vert \phi \vert^2 F \tilde F
\eqn\cpoperator$$
where $\sin(\delta)$ is a measure of CP violation, and $M$
represents the scale of new physics.
Using the anomaly equation, this may be rewritten
$${\cal L}_{CP}={g^2\sin(\delta)\over 12 M^2} \vert \phi \vert^2
\partial_{\mu} j^{\mu~B}.\eqn\rewritten$$

Now during the phase transition, $\phi$ is changing in time near
the bubble walls.  Suppose $\phi$ changes slowly in time,
and can be viewed as nearly homogeneous in space (it is not
hard to modify this analysis for the more realistic case where
spatial gradients are somewhat larger than the time derivatives).
Then we may rewrite the interaction term as
$${\cal L}_{CP}={g^2\sin(\delta)\over 12 M^2}\partial_0 \vert \phi
\vert^2 n_B.$$
If $\phi$ is varying slowly enough, this term is like a chemical
potential for baryon number, $n_B$, and the minimum of the
\REF\sbg{A. Cohen and D. Kaplan, Phys.~Lett. {\bf 199B} (1987) 251;
Nucl.~Phys. {\bf B308} (1988) 913.}
free energy is shifted away from zero, in a manner similar to
the ``spontaneous baryogenesis" picture of Cohen and Kaplan\refmark{\sbg}
An elementary calculation gives for the
location of the new minimum
$$n_B^o={{\cal L}_{CP}={g^2\sin(\delta)\over 48 M^2} \partial_0 \vert \phi
\vert^2}.\eqn\bminimum$$
Since baryon number is violated, the system tries to get to this minimum.
One can easily obtain a rate equation by detailed balance
$${dn_B \over dt}=-18 \Gamma(t) T^{-3}[n_B-n_B^o(t)].\eqn\rateeqn$$
Note that the result is sensible:  the rate vanishes at the minimum;
it is proportional to $\Gamma$ and the number of quark doublets.

Assuming that the time variation of the scalar field is small,
we can make an adiabatic approximation, \ie,
we can assume that everything is in equilibrium except $n_B$ itself.
In this approximation, $\Gamma(t)$ is determined by the instantaneous
value of $\phi$, \ie,
$$\Gamma(t)=\Gamma[\phi(t)]\sim e^{-A m_W(\phi)/(\alpha_W T)}.
\eqn\gammat$$
This turns off for $M_W \sim \alpha_W T$, and can potentially
give a large suppression.  To get a crude estimate of the asymmetry,
take
$$\Gamma= \kappa(\alpha_W T)^4 ~~~M_W < \sigma \alpha_W T$$
$$~~~~~~~~~~~~~~~=0~~~M_W > \sigma \alpha_W T\eqn\crude$$
where $\sigma$ is an unknown numerical constant, about which we will
comment below.
It is now a trivial matter to solve the rate equation, and obtain
$${n_B \over n_{\gamma}} = \alpha_W^6
{3 \sigma^2T^2\sin(\delta ) \kappa \over g^*M^2}.
\eqn\finalasymmetry$$
Here $g^*$ is the number of light degrees of freedom at the phase
transition (of order $100$).  $\alpha_W^6$ is already of order $10^{-9}$,
so to have any chance of obtaining the observed asymmetry in such
a model, one needs $T \sim M$, maximal CP violation, and good luck with
$\kappa$ and $\sigma$.

It is not so obvious how to extend the analysis described above
to the case where  $T \gg M$; on the other hand this is precisely
the situation of interest.  It is also the situation considered
by McLerran \etal\ in their work on the two-Higgs model.\refmark{\mvst}
For general choices of the parameters in the Higgs potential,
a two-Higgs model violates CP.\foot{It is well known that in order to
insure absence of flavor changing neutral currents in such a model,
one must impose a discrete symmetry which also forbids CP violating
terms in the potential.  The authors of ref.~\mvst\ suggest the possibility
that this symmetry is softly broken.  Alternatively, one can simply
view this model as a prototype for more elaborate theories.}
In this theory, the fermions are massless in the unbroken phase,
so one can't simply integrate them out of the zero temperature
theory as before.  McLerran \etal\ argue that one can still integrate
out the fermions at finite temperature.  The point is that in the imaginary
time formalism, fermions have only non-zero discrete frequencies,
$\omega_n = (2n+1) \pi T$.  Thus if one is interested in phenomena
occurring on time and distance scales large compared to $T^{-1}$,
one might expect fermions to be irrelevant.  Indeed, some arguments
supporting this view were given in ref.~\dhs.  If this is correct,
one should compute terms in an effective lagrangian arising from
quark loops.  At lowest order, the top quark (and possibly the b quark)
will be the most important.  As stressed in ref.~\mvst, the possible
operators in this effective lagrangian are not restricted by Lorentz
invariance.  For example, at dimension six, one can write
an operator $\phi_1^* D_o D_i \phi_1 \epsilon_{ijk} F^{jk}$.
These authors argued that this includes $n_{CS} \phi_1^* \partial_o \phi_1$.
Since $\partial_o \phi_1 \ne 0$, this generates a potential for
$n_{CS}$, and therefore for $n_B$.  They
estimated that the resulting baryon asymmetry could easily be consistent
with observations.

Several questions have been raised about this analysis.  First,
just as in the example studied earlier, one must worry that the reaction
shuts down for small $\phi$.  In this model, the problem is potentially
even more severe.  At small $\phi$, one can neglect the quartic terms
in the Higgs potential.  But the quadratic terms in the potential
preserve CP. As a result, one must consider the quartic terms,
leading to suppression by additional powers of $g$.  In terms of the
parameter $\sigma$ we introduced earlier, the estimate of ref.~%
\mvst\ may be written as
$${n_B \over n_\gamma}\sim 10^{-12}~\sigma^4 \times (phases,~couplings).
\eqn\twohiggs$$

Clearly, then, for all of these estimates it is necessary to establish
the size of $\sigma$.  McLerran, and also Shaposhnikov, have argued
that $\sigma$ could be large, perhaps as much as $7$.  This suggestion
follows from the fact
that $\Gamma(\phi)$ is proportional not only to an exponential factor but
to $\phi^7$.  However, the expression in eq.~\fullrate\ cannot be correct
for very small $\phi$, since we know that the rate is non-zero for
$\phi=0$.  Consider, instead, a toy model motivated by
the instanton picture we described earlier.  Take
$$\Gamma = c (\alpha_WT)^{-3.5}\int d \rho \rho^{-8.5}
\exp\left[-{\beta/\alpha_W}(\rho^{-1}+bm^2 \rho)\right]
\eqn\toymodel$$
where
$$m^2=(\alpha_WT)^2+m_W^2(\phi).$$
 The integral over $\rho$
represents the notion that there are trajectories with different scale
for passage over the barrier, and the factor in the exponent represents
the barrier height.  We have added the ``magnetic mass" and
the symmetry breaking mass in quadrature.
The exponents here have been chosen so that the model has the correct
behavior for large and small $\phi$; the rate
decreases monotonically with $\phi$ from its maximum at $\phi=0$.
Typically one finds, evaluating this integral numerically for various
values of the parameters, that $\sigma$ is between 2 and 3.
Of course, this model can't be taken too seriously; what is
really required, as we stressed earlier, are extensive, improved
simulations.

\REF\ckntwo{A. Cohen, D. Kaplan and A. Nelson, Phys.~Lett. {\bf 263B}
(1991) 86.}
Another question has been raised about the analysis of the
two-Higgs model by Cohen, Kaplan and Nelson.\refmark{\ckntwo}
They note that in fact there is no dimension
six operator which is gauge invariant and contains $n_{CS}$.
For example, the operator mentioned above is proportional
to $\vec A \cdot \vec B$, where $\vec B$ is
the magnetic field.  Thus it vanishes at the minima of the
potential.  It is not then clear what biases the baryon violating
transitions.  Similar questions will arise in other models, such
as the models described earlier, once the mass is small compared to the
temperature, if one integrates out the fermions as before.

On the other hand, Cohen \etal\ have performed a different
estimate for this model,\refmark{\ckntwo} which suggests that the asymmetry
which arises may be larger (the connection of the analysis of refs.~%
\ckntwo\ and \mvst\ is not completely clear),
and also raises questions about
the validity of integrating out the fermions.  They noted
that time-varying scalar fields can couple to currents
like hypercharge (or left- or right-handed top
quark number, neutralino number, etc.).  This in turn
induces non-zero densities of the corresponding quantities,
which in turn bias baryon production.  This effect is potentially
important and worth further investigation.

All of the discussion up to now has been based on the assumption
that the wall is thick and slowly moving.  If this is not the
case, the process will not be adiabatic and one might hope
to generate a larger asymmetry.  Indeed, when the wall
is thin and rapidly moving, many effects can come into play.
\REF\turokone{N. Turok and J. Zadrozny, Phys.~Rev.~Lett. {\bf 65}
(1990) 2331.}
Turok and Zadrozny,\refmark{\turok} in their seminal paper on this
subject, considered precisely such a limit.  They argue that
the changing Higgs field itself could force the system
over the barrier through its couplings to Higgs and gauge
fields; CP violation could bias this process.  In this picture, there
is no obvious suppression as $\phi$ grows.  More recently,
\REF\gst{D. Grigoriev, M.E. Shaposhnikov and N. Turok, Imperial
preprint Imperial-TP-92-05 (1992).}
Grigoriev, Shaposhnikov and Turok have presented numerical simulations
in $1+1$ dimension which seem to support this view (somewhat
puzzlingly, these authors find that the asymmetries are largest
at low temperatures).  More work on this aspect of the problem
is needed.

However it is hard to see how this mechanism could
be relevant to phase transitions of the type considered
here.  The energy  deposited by the moving wall per unit volume
is just the internal energy difference between the two
phases.  This is of order
$$\lambda/4 \phi_o^4 \sim 3 \times 10^{-2} T^4\eqn\deltarho$$
for the sorts of Higgs masses which can lead to baryogenesis.
Thus the energy in a volume of order the
sphaleron size, $\pi M_W^{-3}$ is roughly of order $T$, which
is much less than the sphaleron energy.  This suggests that
one must consider biasing of transitions which are already
taking place as the wall approaches.  A simple-minded estimate
suggests that, for a fast-moving wall,
this is likely to be ineffective.\refmark{\dhss}

\REF\cknthree{A. Cohen, D. Kaplan and A. Nelson, UCSD-PTH-91-20 (1991).}
On the other hand, Cohen, Kaplan and Nelson have proposed still
another mechanism which may be relevant to baryon production
in this limit.\refmark{\cknone,\cknthree}
They argue that one should look for processes which can bias
baryon violation in front of the wall, where the rate is unsuppressed.
In particular, they consider mechanisms for generating an asymmetry
in some (approximately conserved) quantum number in front
of the wall.  For example, scattering of top quarks from
the wall, in a theory with CP violation, can lead to an
asymmetry of left vs. right-handed tops in front of the
wall.  A detailed analysis (described at this meeting by
A. Nelson) leads to a substantial asymmetry -- as large as
$10^{-5}$.  The asymmetry tends to become small, however,
if the wall is slow or if it is much thicker than $T^{-1}$.
We have seen that in the simplest models of the phase transition,
the wall tends to be much thicker than this.  It would be of some
interest to construct theories with such thin walls.

\chapter{Conclusions}

{}From all of this effort, we have learned that the baryon asymmetry
may have been created at the electroweak phase transition.
However, before reaching the ideal situation where,
given a model, we can compute the asymmetry,
we need
\item{1.}  better calculations of the B-violating rates
(with and without $\phi$),
\item{2.}  better understanding of the phase transition, and
exploration of models with more strongly first order transitions, and
\item{3.}  better understanding of the processes which generate
the asymmetry.
\endpage
\refout
\figout
\end